\documentclass[twocolumn,showpacs,preprintnumbers,amsmath,amssymb]{revtex4}

\usepackage{graphicx}
\usepackage{dcolumn}
\usepackage{bm}
\usepackage{amsmath}
\usepackage{psfrag}
\usepackage{color}
\usepackage{latexsym,amsfonts}
\usepackage{graphicx}
\providecommand{\href}[2]{#2}   
\definecolor{Blue2}{rgb}{0.,0.,0.8125}
\definecolor{Brown3}{rgb}{0.625,0.25,0.}
\definecolor{Cyan4}{rgb}{0.,0.56,0.56}
\definecolor{Green4}{rgb}{0.,0.56,0.}
\definecolor{LtBlue}{rgb}{0.27,0.42,0.52}
\definecolor{Magenta4}{rgb}{0.5625,0.,0.5625}
\definecolor{Red2}{rgb}{0.8125,0.,0.}

\begin{document}


\title{Electrodynamic Limit in a Model for Charged Solitons}

\author{Manfried Faber}
 \email{faber@kph.tuwien.ac.at}
 
\author{Alexander~P.~Kobushkin}%
\altaffiliation[Permanent address: ]{Bogolyubov Institute for Theoretical Physics,03143, Kiev, Ukraine
        and
        Physical and Technical National University KPI,  
        Prospect Pobedy 37, 03056 Kiev, Ukraine}
 \email{akob@ap3.bitp.kiev.ua}
\affiliation{%
Atominstitut der \"Osterreichischen Universit\"aten\\
        Technische Universit\"at Wien, Wiedner Hauptstr. 8-10\\
        A--1040 Vienna, Austria
}%

\date{\today}

\begin{abstract}
We consider a model of topological solitons where charged particles have finite mass and the electric charge is quantised already at the classical level. In the electrodynamic limit, which physically corresponds to electrodynamics of solitons of zero size, the Lagrangian of this model has two degrees of freedom only and reduces to the Lagrangian of the Maxwell field in dual representation. We derive the equations of motion and discuss their relations with Maxwell's equations. It is shown that Coulomb and Lorentz forces are a consequence of topology. Further, we relate the U(1) gauge invariance of electrodynamics to the geometry of the soliton field, give a general relation for the derivation of the soliton field from the field strength tensor in electrodynamics and use this relation to express homogeneous electric fields in terms of the soliton field.
\end{abstract}

\maketitle

\section{Introduction\label{sec:Introduction}}

The Skyrme model, a low energy approximation of QCD, describes nucleons as soliton configurations of the pion field \cite{Skyrme61,Witten,ANW,MRS}. Such solitons, the skyrmions, appear as classical, stable solutions of non-linear equations of motion derived from a Lagrangian depending only on the pion field, $\vec \pi(x)$, which is a vector in isospin space. Skyrmions posses a non-trivial topological structure.

The intrinsic beauty of the Skyrme model and the well-known success of its application to short-range forces and strongly coupled particles (see, {\it e.g.}, \cite{MRS} and references therein) make it worthwhile to extend its philosophy to the description of long-range forces and electrically coupled particles. An attempt for a realization of such an idea was undertaken in Ref.~\cite{Fab99}.

The most general Lorentz invariant Lagrangian including only time derivatives to second power reads
\begin{equation}\label{GenaralLagrangian}
\mathcal{L}=A\mathrm{Tr}\left\{L_\mu L^\mu\right\} + B\mathrm{Tr} \left\{ \left[ L_\mu,L_\nu \right] \left[ L^\mu,L^\nu \right] \right\},
\end{equation}
where $L_\mu$ is the left current 
\begin{equation}\label{LeftCurrent}
L_\mu=\partial_\mu U U^\dagger
\end{equation}
with the $2\times 2$ chiral field
\begin{equation}\label{U-field}
U=\exp\left(i\vec{\tau}\vec{\pi}(x)\right),
\end{equation}
$\vec{\tau}$ are the isospin Pauli matrices. $A$ and $B$ are some constants specified by appropriate physical requirements.  

In the Skyrme model $A$ and $B$ are given by
\begin{equation}\label{SkyrmeConstants}
A=-\frac{F^2_\pi}{16} \quad \mathrm{and} \quad B=\frac1{32e^2_S},
\end{equation}
where $F_\pi=186$~MeV is the pion decay constant and $e_S$ is related to static properties of the nucleon and the $\Delta$-resonance. The terms proportional to $A$ and  $B$ are called ``kinetic term'' and ``Skyrme term'', respectively.

A static hedgehog ansatz for the chiral field is usually assumed
\begin{equation}\label{HedgehogGeneral}
\vec{\pi}(\mathbf{x})= \Theta(r) \, \vec n,\quad \vec n=\frac{\mathbf{r}}{r}, \quad r=|\mathbf{r}|.
\end{equation}
The profile function $\Theta(r)$ is determined from the minimum of the classical energy
\begin{equation}\label{StaticEnergy}
\begin{split}
E_\mathrm{cl}
=&4\pi\int_0^\infty dr r^2 \left\{
8A\left[\Theta'^2 + 2\frac{\sin^2\Theta}{r^2}\right]+ \right.\\
&+\left.
16B\frac{\sin \Theta}{r^2}\left[
\frac{\sin \Theta}{r^2}+2\Theta'^2
\right]
\right\}.
\end{split}
\end{equation}
The boundary conditions 
\begin{eqnarray}
 \lim_{r \to 0} \Theta(r)&=&n\pi  \quad \mathrm{with\ integer}  \quad n, \label{BCatOrigion}\\
\lim_{r \to \infty} \Theta(r)&=&0 \label{BCatInftyS}
\end{eqnarray}
follow from the requirement to have finite energy $E_\mathrm{cl}$. The number of coverings $n$ of the internal sphere $\mathrm S^3$, the topological charge, is interpreted as baryon number. Such solitons can be regarded as fermions \cite{MRS}. The boundary condition (\ref{BCatInftyS}) reduces the chiral field $U$ at space-like infinity to a trivial configuration
\begin{equation}\label{Skyrme_BC}
  \lim_{r \to \infty}U(\vec \pi(x))=1.
\end{equation}
Due to the boundary condition (\ref{BCatInftyS}) and (\ref{Skyrme_BC}) the  profile function $\Theta(r)$ decreases as $r^{-2}$ and the energy density like $r^{-6}$ at large distances. Therefore, the Skyrme model cannot describe Coulombic interactions between solitons, forces behaving like $r^{-2}$. For the description of such forces a different behaviour of the profile function $\Theta(r)$ at infinity is necessary
\begin{equation}\label{BC_new}
\lim_{r \to \infty} \Theta(r)=\mathrm{const}\neq 0.
\end{equation}
With this choice the chiral field $\vec \pi(x)$ depends on the direction in space,  $\lim_{r\to \infty}\vec \pi(x)=\vec n(x) \cdot\mathrm{const}$, and has two independent components only.

For the boundary condition (\ref{BC_new}) the classical energy (\ref{StaticEnergy}) is divergent due to the term proportional to $A$. Finite energies can be obtained only, setting $A=0$.

To get stable solitonic solutions one has to fulfill the Hobart-Derrick \cite{HD64} necessity condition. The Lagrangian has to include a  compressing and a dissolving term. The kinetic term with its two derivatives is compressing and the Skyrme term with four derivatives is dissolving solitons. Since the kinetic term in the Lagrangian (\ref{GenaralLagrangian}) is forbidden for the stabilisation of electrically charged solitons there is only the possibility to stabilize solitons by a term without derivatives, i.e. a potential term which also tends to shrink solitons. A Lagrangian with such a mechanism was recently suggested in Ref.~\cite{Fab99}.

In Section~\ref{SecMTF} we give a short review of this model. This model admits solitons with integer multiples of the elementary electric charge $e_0$ only. Charged solitons, similar to skyrmions, fulfil Pauli's exclusion principle and thus possess the properties of topological fermions \cite{Fab99}.

It is natural to ask the following question: How is this model of ``effective electrodynamics'' (later on we will call it the model of topological fermions, MTF) related to Maxwell's electrodynamics and to electrodynamics extended by Dirac monopoles \cite{Di48}? To answer this question we introduce in Section~\ref{sec:edlimit} an idealized model which physically corresponds to point-like charges, but keeps the main topological properties of the MTF. This limit can be reached by putting the natural length scale $r_0$ of the model to zero. We call this limit the electrodynamic limit of charged solitons.

In Section~\ref{SecEOM} we  reformulate the MTF equations of motion in terms of electric and magnetic field strengths and compare them with Maxwell's equations. We show that the inhomogeneous Maxwell's equations are a consequence of the topological obstruction of the colour field. The main difference to Maxwell's theory is in the types of charges which can be described. In Maxwell's theory any charge distribution is possible. In the solitonic description only integer multiples of the elementary charge can appear. The second pair of equations follows from the equations of motion. They correspond to the homogeneous Maxwell equations extended by a magnetic current. The appearance of magnetic currents is a result of the non-Abelian nature of the colour field.

The magnetic currents coming from the soliton description are different form the Dirac magnetic currents, because they propagate in the vacuum with the speed of light and cannot be associated with massive magnetic charges, see Section~\ref{Propagating}. In this section we also investigate scattering of electromagnetic waves in the background of solitons.

We derive Coulomb and Lorentz forces in Section~\ref{sec:Forces}. In Section~\ref{sec:gauge} we relate the U(1) gauge invariance of electrodynamics to the geometry of the soliton field, give a general relation for the derivation of the soliton field from the field strength tensor in electrodynamics and use this relation in Section~\ref{sec:homogeneous} to express homogeneous electric fields in terms of the soliton field.

\begin{figure*}[htb]
  \centering
  \psfrag{S3}{\textcolor{Cyan4}{$\mathrm{S}^3$}}
  \psfrag{S2}{\textcolor{Magenta4}{$\mathrm{S}^2_\mathrm{equ}$}}
  \psfrag{u0}{\textcolor{Blue2}{$q_0$}}
  \psfrag{u1}{$q_1$}
  \psfrag{u2}{$q_2,q_3$}
  \psfrag{Hp}{$\Lambda(q_0)$}
  \includegraphics[width=0.65\textwidth]{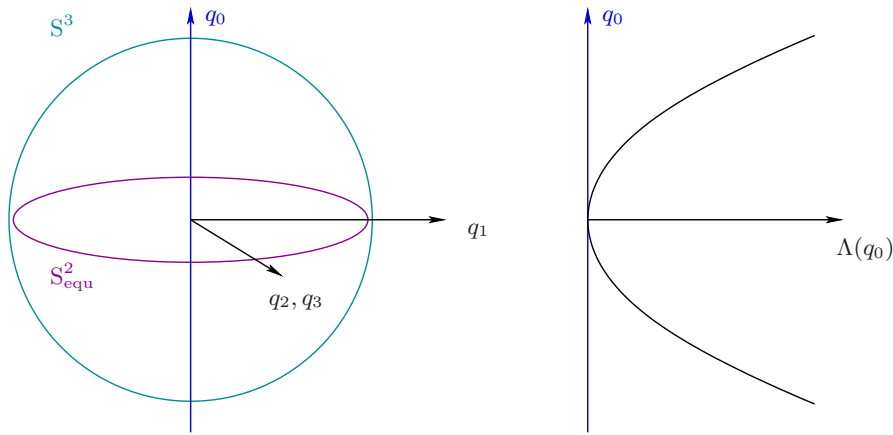}
\caption{Shape of the potential term $\Lambda(q_0)$ leading to spontaneous symmetry breaking in the equatorial $\mathrm{S}^2_\mathrm{equ}$ of $\mathrm{S}^3$.}
\label{figpot}
\end{figure*}

\section{Model Lagrangian}\label{SecMTF}

A Lagrangian suited for the description of charged solitons was discussed in Ref.~\cite{Fab99}. Here, we give only a short review of this model. The manifold of the soliton field $Q$ of the MTF is SO(3), it has therefore three degrees of freedom, which can be   encoded in an SU(2) field,
\begin{equation}\label{ExpRep}
Q(x)=e^{i\vec \alpha(x) \vec \sigma}=\cos \alpha(x) + i \vec n(x)\vec \sigma \sin \alpha(x),
\end{equation}
where the field  $\vec \alpha(x)$ depends on the position $x^\mu=(\mathbf{r},t)$ in Minkowski space-time, $\vec{\sigma}$ are the Pauli matrices \footnote{We use the summation convention that any capital Latin index that is repeated in a product is automatically summed on from 1 to 3. The arrows on variables in the internal ``colour'' space indicate the set of 3 elements $\vec{q}=(q_1, q_2, q_3)$ or $\vec{\sigma}=(\sigma_1, \sigma_2, \sigma_3)$ and $\vec{q} \vec{\sigma}= q_K \sigma_K$. We use the wedge symbol $\wedge$ for the external product between colour vectors $(\vec{q}\wedge\vec{\sigma})_A=\epsilon_{ABC}q_B \sigma_C$. For the components of vectors in physical space $\mathbf{x}=(x,y,z)$ we employ small Latin indices, $i,j,k$ and a summation convention over doubled indices, {\it e.g.} $(\mathbf{E} \times \mathbf{B})_i = \epsilon_{ijk} E_j B_k$. Further we use the metric $\eta = \mathrm{diag}(1,-1,-1,-1)$ in Minkowski space.} and
\begin{equation}
\alpha(x)=|\vec \alpha(x)|,  \quad
\vec n(x)=\frac{\vec{\alpha}(x)}{\alpha(x)}.
\end{equation}
Later on we will also use an equivalent parametrization
\begin{equation}\label{Q(x)}
Q(x)=q_0(x)+i\sigma_K q_K(x) \quad \text{with} \quad q_0^2+\vec{q}\hspace{0.4mm}^2=1,
\end{equation}
where
\begin{equation}\label{qRep}
q_0(x)=\cos \alpha(x),  \quad \vec q(x) = \vec n(x) \sin \alpha(x).
\end{equation}

To specify the MTF Lagrangian we define in analogy to Eq.~(\ref{LeftCurrent}) the connection $\vec{\Gamma}_\mu$ by
\begin{equation}\label{Gamma}
\vec{\Gamma}_\mu = \frac{1}{2i} \mathrm{Tr} (\vec{\sigma}\partial_\mu Q Q^\dagger)
\end{equation}
and the curvature tensor
\begin{equation}\label{intcurvature1}
\vec{R}^{\mu \nu} = \partial^\nu \vec{\Gamma}^\mu - \partial^\mu \vec{\Gamma}^\nu - \vec{\Gamma}^\mu \wedge \vec{\Gamma}^\nu.
\end{equation}
In the Appendix we discuss the relation of the connection $\vec{\Gamma}^\mu(x)$ and the curvature  $\vec{R}^{\mu \nu}$ to geometry. The three-component colour vectors (\ref{Gamma}) and (\ref{intcurvature1}) are the components of su(2)-algebra elements.

The Lagrangian of the MTF reads
\begin{eqnarray}\label{Lagrangian}
{\cal L} = - \frac{\alpha_f \hbar c}{4 \pi}
\left(\frac{1}{4} \; \vec{R}_{\mu \nu} \vec{R}^{\mu \nu} 
+\Lambda(q_0)\right)
\end{eqnarray}
with Sommerfeld's fine-structure constant $\alpha_f = \frac{e_0^2}{4 \pi \varepsilon_0 \hbar c }$.
The potential term $\Lambda(q_0)$ is a function of the trace Tr$Q=2q_0$, as shown in Fig.~\ref{figpot}. It is zero for the equatorial $\mathrm{S}^2_\mathrm{equ}$  ($q_0=0$), monotonically increasing with $|q_0|$. Increasing with an even power of $q_0$ the potential term reads
\begin{eqnarray}\label{potential}
\Lambda(q_0)=\frac{1}{r^4_0}\left(\frac{\mathrm{Tr}Q}{2}\right)^{2m}&=&
       \frac{1}{r^4_0}\cos^{2m}\alpha(x),  \nonumber \\ 
&&m=1,2,3,\dots \ ,
\end{eqnarray}
where $r_0$ is a dimensional parameter of the MTF. It should be mentioned that the curvature term in the Lagrangian (\ref{Lagrangian}) of the MTF is proportional to the Skyrme term, see Eqs.~(\ref{GenaralLagrangian}) and (\ref{SkyrmeConstants})
\begin{equation}\label{Skyrme}
\frac{1}{32} {\rm Tr} \left\{ \left[ L_\mu,L_\nu \right] \left[ L^\mu,L^\nu \right] \right\} = - \frac{1}{4} \; \vec{R}_{\mu \nu} \vec{R}^{\mu \nu}.
\end{equation}

This curvature term has the shape of a Yang-Mills Lagrangian for a gauge field $\vec\Gamma_\mu$ and a field strength tensor $\vec R_{\mu\nu}$ defined by (\ref{intcurvature1}). There is the distinction to the connection field (\ref{Gamma}) that in a Yang-Mills theory the gauge field does not follow the Maurer-Cartan equation (\ref{maurercartan}) and has therefore more degrees of freedom. It is well-known that in the vacuum two of these degrees of freedom are gauge degrees of freedom and only two degrees are physical. It is interesting to note that in a gauge theory a term like the ``kinetic term'' of the Skyrme model would be a mass term for the gauge field $\vec{\Gamma}^\mu(x)$. This underlines that such a term is not allowed in a model for massless photons.

From the Lagrangian (\ref{Lagrangian}) one can derive the equation of motion
\begin{equation}\label{GeneralEOM}
\partial_\mu [ \vec{\Gamma}_\nu \times \vec{R}^{\mu \nu} ] \, + \, \vec{q} \, \frac{d \Lambda}{d q_0} \, = 0 .
\end{equation}
This equation must be supplied with two boundary conditions. The boundary condition at infinity is derived from general principles of soliton models \cite{Raja}: As $r \to \infty$ the localized solution, if any, must approach field values at one of the absolute minima of the potential $\Lambda(q_0)$. According to its construction $\Lambda(q_0)$ has a minimum on the equatorial $\mathrm{S}^2_\mathrm{equ}$ of $\mathrm{S}^3$ in Fig.~\ref{figpot}. This means that $\alpha(\infty)$ must be a half-integer multiple of $\pi$
\begin{equation}\label{Boun-cond_infty}
\alpha(\infty)=\pi\left(\frac12+n_\infty\right), \qquad \text{with integer $n_\infty$},
\end{equation}
which is equivalent to
\begin{equation}\label{Faber_BC}
\lim_{r \to\infty}Q(x)=i\vec \sigma\vec n(x). 
\end{equation}
This is nothing more than a non-singular mapping
\begin{equation}\label{mapping}
\mathrm{S}^2_\mathrm{phy} \to \mathrm{S}^2_\mathrm{equ}.
\end{equation}
Such mappings can be classified by homotopy classes which are characterised by an integer $Z$, $\pi_2(\mathrm{S}^2)=Z$.

For solitons of finite energy it is allowed to deviate from the vacuum value (\ref{Faber_BC}) in some finite region of space only. For a single soliton at the origin the boundary condition comes from the same arguments as in the Skyrme model and reads
\begin{equation}\label{Boun-cond_0}
\alpha(0)=n_0\pi, \qquad \text{with integer } n_0.
\end{equation} 
Without loss of generality we set $n_0=0$.

The essential difference is that in the Skyrme model the $U$-field maps a soliton to complete internal $\mathrm S^3$, $\mathrm R^3\to \mathrm S^3$. But the $Q$-field maps the soliton onto half of $\mathrm S^3$, $\mathrm R^3 \to $SO(3). Finally, the solitons of the MTF are characterized by two integers, $n_w=n_\infty-2 n_0$ and $Z$. In this paper we do not discuss $n_w$ because it is of no importance in the limit of point-like solitons, which is the main subject of the present work. $n_w$ was already studied in Ref.~\cite{Fab99} where the model was introduced.  $Z$ has the physical meaning of electric charge as will be discussed in Section~\ref{sec:edlimit}.

\begin{figure}
\psfrag{alpha(r)}{$\alpha(r)$}
\psfrag{-p2}{$-\frac{\pi}{2}$}
\psfrag{p2}{$+\frac{\pi}{2}$}
\psfrag{r/r_0}{$r/r_0$}
\psfrag{Z=1}{$Z=1$}
\psfrag{Z=-1}{$Z=-1$}
\psfrag{0}{$0$}
\psfrag{10}{$10$}
\psfrag{2}{$2$}
\psfrag{4}{$4$}
\psfrag{6}{$6$}
\psfrag{8}{$8$}
\centerline{\includegraphics[width=0.5\textwidth]{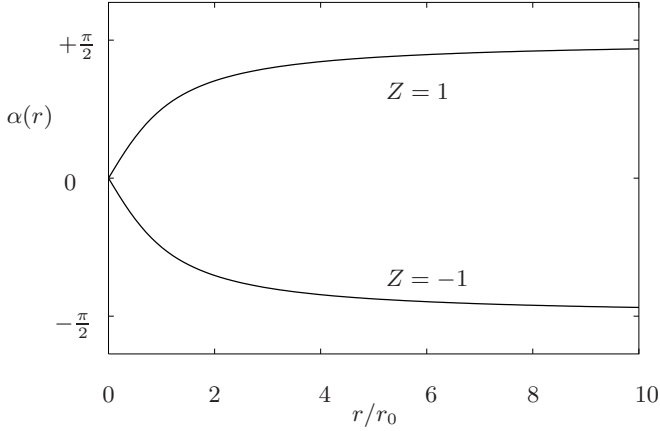}}
\caption{Two profile functions $\alpha(r)$ for $m=3$. According to the hedgehog form (\ref{hedgehog}), (\ref{profile}) and (\ref{Qelect}) positive (negative) angles describe positive (negative) charges.
}
\label{fig:alpha}
\end{figure}

The lowest non-trivial solution for the equation of motion (\ref{GeneralEOM}) we get for the hedgehog ansatz
\begin{equation}\label{hedgehog}
\quad \alpha = \alpha(r), \quad \vec n  = \frac{\mathbf r}{r}
\end{equation}
for the soliton field $Q(x)$. For this ansatz the equations of motion reduce to the non-linear second order differential equation
\begin{equation}\label{nlDE}
\partial^2_r \cos \alpha \; + \; \frac{(1 - \cos^2 \alpha) \cos \alpha}{r^2} \; - \; \frac{r^2}{2} \frac{d \Lambda}{d q_0}\; = \; 0.
\end{equation}
The solution of lowest energy for the power $m=3$ in the potential term (\ref{potential}) was given in \cite{Fab99} in analytical form, see also Fig.~\ref{fig:alpha},
\begin{equation}\label{profile}
\alpha(r) \; = \; \text{atan}(\rho) \quad \text{with} \quad \rho = \frac{r}{r_0}
\end{equation}
and the total energy for a single soliton reads
\begin{equation}
E_1 \; = \; \frac{\alpha_f \hbar c}{r_0} \frac{\pi}{4} \quad \text{with} \quad \alpha_f \hbar c = 1.44 \text{~MeV~fm} .
\end{equation}
If we compare this expression with the energy of the electron of $m_e c^2=$0.511 MeV, we can fix $r_0 = 2.21$\,fm, a value which is very close to the classical electron radius $\alpha_f \hbar/m_e c=2.82$\,fm. This comparison to the electron leads to the interpretation of the connection field $\vec\Gamma_\mu$ as a dual non-Abelian vector field in natural units 
\begin{equation}\label{Cgeneral}
\vec C_\mu =  - \frac{e_0}{4 \pi \varepsilon_0} \vec \Gamma_\mu
\end{equation}
and the curvature as a dual non-Abelian field strength tensor
\begin{equation}\label{dualfield}
\hspace{0.8mm}{^*}\hspace{-0.8mm}\vec F_{\mu \nu} = - \frac{e_0}{4 \pi \varepsilon_0 c} \vec R_{\mu \nu} = 
\left( \begin{array}{cccc}
 0 & \vec B_x & \vec B_y & \vec B_z \\
 -\vec B_x & 0 & \vec E_z/c & -\vec E_y/c \\
 -\vec B_y & -\vec E_z/c & 0 & \vec E_x/c \\
 -\vec B_z & \vec E_y/c & -\vec E_x/c & 0 \\
\end{array} \right) .
\end{equation}

For the static soliton (\ref{hedgehog}) the vector field $\vec C_\mu$ reads \cite{Fab99}
\begin{equation}\label{CHedgeHog}
\begin{split}
&\vec C_t = 0, \quad
\vec C_r = - \frac{e_0}{4\pi \varepsilon_0} \alpha^\prime(r) \, \vec n,\\
&\vec C_\vartheta = - \frac{e_0}{4\pi \varepsilon_0} \frac{\sin \alpha}{r} \left( \cos \alpha \, \vec{e}_\theta - \sin \alpha \, \vec{e}_\phi \right),\\
&\vec C_\varphi = - \frac{e_0}{4\pi \varepsilon_0} \frac{\sin \alpha}{r} \left( \sin \alpha \, \vec{e}_\theta + \cos \alpha \,  \vec{e}_\phi \right),
\end{split}
\end{equation}
the magnetic field is vanishing and the electric field strength reads
\begin{equation}\label{EFieldStrengthHedgeHog}
\begin{split}
&\vec E_r = - \frac{e_0}{4\pi \varepsilon_0} \frac{\sin^2 \alpha}{r^2} \, \vec n, \\
&\vec E_\vartheta = - \frac{e_0}{4\pi \varepsilon_0} \; \frac{\alpha^\prime(r) \, \sin \alpha}{r} \, (\cos \alpha \; \vec{e}_\theta - \sin \alpha \; \vec{e}_\phi),\\
&\vec E_\varphi = - \frac{e_0}{4\pi \varepsilon_0}\; \frac{\alpha^\prime(r) \, \sin \alpha}{r} \, (\sin \alpha \; \vec{e}_\theta + \cos \alpha \;  \vec{e}_\phi) .
\end{split}
\end{equation}
where $\vec n$, $\vec{e}_\theta$ and $\vec{e}_\phi$ are spherical unit vectors in internal colour space corresponding to unit vectors $\mathbf e_r$, $\mathbf e_\vartheta$ and $\mathbf e_\varphi$ in coordinate space.

According to (\ref{profile}) $\alpha^\prime(r)$ behaves like $1/r^2$, therefore $\vec E_\vartheta$ and $\vec E_\varphi$ decay faster than $\vec E_r$ which for distances much larger than $r_0$ behaves like $1/r^2$ and points in colour $\vec n$-direction. In this limit the electric field strength reads
\begin{equation}\label{El_hedgehog}
\vec{\mathbf{E}}=-\frac{e_0\mathbf{r}}{4\pi \varepsilon_0 r^3}\vec n.
\end{equation}
This shows that in the limit $r_0 \to 0$, we call it the electrodynamic limit, the soliton field reduces to a one-component Abelian field. The investigation of this limit is the main subject of this paper.

\section{Electrodynamic limit}\label{sec:edlimit}

One of the main features of the soliton model of electro-magnetically coupled particles is related to the appearance of topologically non-trivial field configurations. At regions outside of soliton cores the potential term (\ref{potential}), see also Fig.~\ref{figpot}, restricts the values of the soliton field $Q(x)$ to $q_0=0 \Leftrightarrow \vec q = \vec n$, field values which form a two-dimensional sphere $\mathrm{S}^2_\mathrm{equ}$ with
\begin{equation}\label{Qlimit}
Q=i\vec\sigma \vec n, \quad \alpha=\frac\pi2 .
\end{equation}
Although for the hedgehog solution (\ref{hedgehog}) the $\vec{n}$-field is singular at the soliton center, due to the smooth profile function (\ref{profile}) suppressing the singularity at the origin the $Q$-field is everywhere regular. In this article we do not want to study the short distance behaviour of solitons, we would like to concentrate on the electrodynamic properties of the soliton field and compare them with the well-known models of electrodynamics. For this aim we consider the limit where $r_0$, the length scale of the model, defined in Eq.~(\ref{potential}), is sent to zero, $r_0 \to 0$, which keeps the topological properties of the MTF, but formally correspond to point-like charges. In this limit, we call it the electrodynamic limit, these non-trivial field configurations are therefore described by the $\vec n$-field only and solitons shrink to dual Dirac monopoles, to point-like charges. The $Q$-field reduces to its asymptotic value (\ref{Qlimit}) everywhere in space-time besides the world-lines of electric charges.

Inserting (\ref{Qlimit}) in (\ref{Gamma}) and (\ref{intcurvature1}) the connection $\vec{\Gamma}^\mu(x)$ and the curvature $\vec{R}^{\mu \nu}(x)$ read
\begin{equation}\label{At_infinity}
\begin{split}
\vec{\Gamma}^\mu(x)&=\partial^\mu \vec{n}(x) \wedge \vec{n}(x),\\
\vec{R}^{\mu \nu}(x)&=\partial^\mu \vec{n}(x) \wedge \partial^\nu \vec{n}(x).
\end{split}
\end{equation}
$\vec R_{\mu\nu}$ points in $\vec n$-direction. From its modulus one can construct an Abelian field strength tensor and identify it with the dual electromagnetic field strength in international (SI) units
\begin{eqnarray}\label{AbelianFS}
{\hspace{1mm}^*}\hspace{-1mm}f_{\mu \nu}(x) &=&  -\frac{e_0}{4 \pi \varepsilon_0 c} \vec{R}^{\mu\nu}\vec{n} =\\
&=& - \frac{e_0}{4 \pi \varepsilon_0 c} [ \partial_\mu \vec{n}(x) \wedge \partial_\nu \vec{n}(x) ] \vec{n}(x)
\nonumber \\
&=&\left( \begin{array}{cccc}
 0 & B_x & B_y & B_z \\
 -B_x & 0 & \frac{E_z}{c} & \frac{-E_y}{c} \\
 -B_y & \frac{-E_z}{c} & 0 & \frac{E_x}{c} \\
 -B_z & \frac{E_y}{c} & \frac{-E_x}{c} & 0 \\
\end{array} \right).
\end{eqnarray}
The Abelian field strength $f_{\mu \nu}(x)$ we get from its dual by $f_{\mu \nu}(x) = -\frac{1}{2} \epsilon_{\mu\nu\rho\sigma} {\hspace{0.3mm}^*}\hspace{-0.3mm}f^{\rho\sigma}(x)$ with $\epsilon^{0123}=1$.

The Lagrangian of the MTF (\ref{Lagrangian}) reduces in the electrodynamic limit to
\begin{equation}\label{EDLagrangian}
\begin{split}
{\cal L_{\rm ED}} &=  - \frac{\alpha_f \hbar c}{4 \pi} \frac{1}{4} \; (\vec{n}\vec{R}_{\mu \nu}) (\vec{n}\vec{R}^{\mu \nu})=\\
&= - \frac{1}{4 \mu_0} {\hspace{0.5mm}^*}\hspace{-0.5mm}f_{\mu \nu}(x){\hspace{0.5mm}^*}\hspace{-0.5mm}f^{\mu \nu}(x),
\end{split}
\end{equation}
with the dual Abelian field strength tensor $^{*}\hspace{-0.5mm}f^{\mu \nu}(x)$ given by (\ref{AbelianFS}).

By Gau\ss's law the electric charge $Q_\mathrm{el}({\cal S})$ enclosed by a two-dimensional surface $\mathcal{S}$ is related to the flux through this surface
\begin{equation}\label{Gausslaw}
Q_\mathrm{el}(\mathcal{S})=\int_V d^3x \rho(x)=c\varepsilon_0\oint_{{\cal S}(u,v)} du dv \, ^*f_{u v},
\end{equation}
where $V$ is the volume enclosed by the surface $\mathcal S$. Here we parametrise the surface with parameters $u$ and $v$ and $\rho(x)$ is the electric charge density. Substituting in the r.h.s. of this expression the dual electromagnetic field strength (\ref{AbelianFS}) the expression  (\ref{Gausslaw}) reads
\begin{equation}\label{surfint}
Q_\mathrm{el}(\mathcal{S})=-\frac{e_0}{4\pi}
\oint_{{\cal S}(u,v)} du dv \, [ \partial_u \vec{n} \wedge \partial_v \vec{n} ] \vec{n}.
\end{equation}
The integral at the r.h.s. of (\ref{surfint}) coincides with the topological (winding) number $Z({\cal S})$ for the mapping \cite{Raja}
\begin{equation}
\mathcal{S} \to \mathrm{S}^2_\mathrm{equ}
\end{equation}
which reads 
\begin{equation}\label{topologicalnumber}
 Z({\cal S})=\frac1{4\pi}\oint_{{\cal S}(u,v)} du dv \, [ \partial_u \vec{n} \wedge \partial_v \vec{n} ] \vec{n}.
\end{equation}
Therefore, the electric charge enclosed by $\mathcal S$ is given by
\begin{equation}\label{Qelect}
Q_\mathrm{el}({\cal S})=-Z({\cal S})e_0.
\end{equation}

The soliton solution (\ref{hedgehog}) and (\ref{profile}) can be interpreted as a model for an elementary electric charge with its mass located in some region of space around the origin. Further on, we call such charges for simplicity electrons or positrons depending on their sign.

Since the Lagrangian is Lorentz invariant it is obvious that electrons experience Lorentz contraction and the relativistic increase of mass with velocity. Nevertheless, it is nice to see microscopically how this relativistic properties appear at the level of the soliton field. Lorentz contraction appears as a local increase of the curvature of the soliton field, necessary to accelerate the soliton field in front of a moving particle. Therefore, electric and kinetic (magnetic) energies increase with the velocity of an electron. This together with the decreasing potential energy leads to a relativistic increase of the total energy \cite{Fab99}. The velocity dependence of mass is therefore understood as a velocity dependence of the energy content of the soliton. Electron-positron creation and annihilation are possible already at a classical level conserving topological quantum numbers.

\section{Equations of motion in the electrodynamic limit and relation to Maxwell-equations\label{SecEOM}}

There are two types of electric monopoles. Negatively charged monopoles with positive coverings of $\mathrm{S}^2_\mathrm{equ}$, electrons, those of opposite charge, positrons. The world-lines of negative and positive monopoles can join, this corresponds to charge cancellation or separation. For $N$ separate world-lines of particles or antiparticles as functions of their eigentime $\tau_i$
\begin{equation}\label{worldline}
X^\mu(\tau_i) \qquad i=1,\cdots,N
\end{equation}
we can define a current vector
\begin{equation}\label{currvec}
j^\mu=-e_0 c \sum_{i=1}^N \int d\tau_i \frac{dX^\mu(\tau_i)}{d\tau_i} \, \delta^4 (x-X(\tau_i))
= (c \rho, \mathbf{j}).
\end{equation}
We can now use (\ref{Gausslaw}) and apply Gau\ss's law
\begin{eqnarray}\label{intinhmax}
\frac{1}{2}&&\hspace{-6mm} \oint_{\cal S} dx^\mu dx^\nu {^*f}_{\mu\nu}= \nonumber \\
&=&\frac{1}{6} \int_V dx^\mu dx^\nu dx^\rho 
(\partial_\mu {\hspace{0.5mm}^*}\hspace{-0.5mm}f_{\nu\rho}+\partial_\nu {\hspace{0.5mm}^*}\hspace{-0.5mm}f_{\rho\mu}+\partial_\rho {\hspace{0.5mm}^*}\hspace{-0.5mm}f_{\mu\nu})= \nonumber\\
&=&\frac{1}{6} \int_V dx^\mu dx^\nu dx^\rho \epsilon_{\mu\nu\rho\sigma} \partial_\lambda f^{\lambda\sigma}=\\
&=&\frac{\mu_0}{6} \int_V dx^\mu dx^\nu dx^\rho \epsilon_{\mu\nu\rho\sigma} j^\sigma. \nonumber
\end{eqnarray}
Since the three-dimensional integration volume is arbitrary we get the inhomogeneous Maxwell-equations
\begin{equation}\label{inhmax}
\partial_\mu f^{\mu\nu}=\mu_0 j^\nu
\quad \Leftrightarrow \quad
\begin{cases}
&\frac{\rho}{\varepsilon_0} = \nabla \mathbf{E},\\
&\frac{\mathbf{j}}{\varepsilon_0} = c^2 \nabla \times \mathbf{B} - \partial_t\mathbf{E}.
\end{cases}
\end{equation}
As shown above these equations are a consequence of the topological obstruction of the $\vec{n}$-field.

In the following we derive the equations of motion and discuss their relation to the homogeneous Maxwell-equations. According to (\ref{AbelianFS}) and (\ref{EDLagrangian}) in the limit of electrodynamics the action $S$ is a functional of $\vec{n}(x)$ only and is given by
\begin{eqnarray}\label{eldynaction}
S[\vec{n}]&=&\int d^4x{\cal L_{\rm ED}}=
 \\
&=&- \frac{\alpha_f \hbar c}{4 \pi} \frac{1}{4} \int d^4x (\partial_\mu\vec{n} \wedge \partial_\nu\vec{n})(\partial^\mu\vec{n} \wedge \partial^\nu\vec{n}).\nonumber
\end{eqnarray}
The equations of motion we get by a variation of $\vec{n}(x)$ under the constraint $\vec{n}^2=1$. This variation we perform with an $x$-dependent rotation of $\vec{n}(x)$ by an angle $\epsilon \vec{\zeta}(x)$ with an arbitrary vector field $\vec{\zeta}(x)$
\begin{equation}\label{vary}
\vec n \to \vec n^\prime = e^{i\epsilon\vec{\zeta}\vec{T}}\vec{n},
\end{equation}
where $\vec{T}$ is the generator of vector rotations $(T_A)_{BC}=-i\epsilon_{ABC}$\footnote{The generators $T_A$ in the adjoint representation of SU(2) obey $[T_A,T_B]=i \epsilon_{ABC} T_C$ and Tr$(T_A T_B)=2\delta_{AB}$.}. We expect the equations of motion from the component of $\vec{\zeta}$ transversal to $\vec{n}$
\begin{equation}\label{transv}
\vec \zeta \vec n = 0.
\end{equation}
The parallel component of $\vec{\zeta}$ leads to a trivial relation only. A necessary condition for an extremum of the action is the vanishing of the first functional derivative of $S[\vec{n}]$ which is defined by
\begin{equation}\label{variation}
\lim_{\epsilon \to 0} \frac{S[e^{i\epsilon\vec{\zeta}\vec{T}}\vec{n}]-S[\vec{n}]}{\epsilon}=
\int d^4x\;\vec{\zeta}(x)\,\frac{\delta S[e^{i\vec{\zeta}\vec{T}}\vec{n}]}{\delta\vec{\zeta}(x)}=0.
\end{equation}

Expanding up to the first power in $\epsilon$
\begin{equation}\label{elemvar}
\begin{split}
&e^{i\epsilon\vec{\zeta}\vec{T}}\vec{n}=
\vec{n}+i\epsilon(\vec{\zeta}\,\vec{T})\vec{n}=
\vec{n}+\epsilon\vec{n}\wedge\vec{\zeta},\\
&\partial_\mu(e^{i\epsilon\vec{\zeta}\vec{T}}\vec{n})=
\partial_\mu\vec{n}+\epsilon(\vec{n}\wedge\partial_\mu\vec{\zeta}+\partial_\mu\vec{n}\wedge\vec{\zeta}),\\
&S[e^{i\epsilon\vec{\zeta}\vec{T}}\vec{n}] \propto
-\frac{1}{4} \int d^4x(\partial^\mu\vec{n}\wedge\partial^\nu\vec{n})\\
&\hspace{1cm}\left\{
[\partial_\mu\vec{n}+4\epsilon(\vec{n}\wedge\partial_\mu\vec{\zeta}+\partial_\mu\vec{n}\wedge\vec{\zeta})]\wedge\partial_\nu\vec{n}
\right\}
\end{split}
\end{equation}
we get from (\ref{variation})
\begin{equation}
0=\int d^4x
[(\vec{n}\wedge\partial_\mu\vec{\zeta}+\partial_\mu\vec{n}\wedge\vec{\zeta})\wedge\partial_\nu\vec{n}]
 (\partial^\mu\vec{n}\wedge\partial^\nu\vec{n}).
\end{equation}
Using
\begin{equation}\label{varextprod}
\begin{split}
&(\vec{n}\wedge\partial_\mu\vec{\zeta})\wedge\partial_\nu\vec{n}=
-\vec{n}(\partial_\mu\vec{\zeta}\;\partial_\nu\vec{n}),\\
&(\partial_\mu\vec{n}\wedge\vec{\zeta})\wedge\partial_\nu\vec{n}=
\vec{\zeta}(\partial_\mu\vec{n}\;\partial_\nu\vec{n})-
\partial_\mu\vec{n}(\vec{\zeta}\;\partial_\nu\vec{n})
\end{split}
\end{equation}
it follows
\begin{equation}\label{doublecross}
0=\int d^4x
[(\partial_\mu\vec{\zeta}\;\partial_\nu\vec{n})\vec{n}-
(\partial_\mu\vec{n}\;\partial_\nu\vec{n})\vec{\zeta}\,]
 (\partial^\mu\vec{n}\wedge\partial^\nu\vec{n}).
\end{equation}
Since contributions antisymmetric in $\mu$ and $\nu$ vanish, the term proportional to $(\partial_\mu\vec{n}\;\partial_\nu\vec{n})$ does not contribute. From the other term we get by partial integration of $\partial_\mu\vec{\zeta}$ again a vanishing antisymmetric term and
\begin{equation}\label{eqofmotionhommax}
0=\int d^4x
(\vec{\zeta}\;\partial_\mu\vec{n})\;\partial_\nu[\vec{n}(\partial^\mu\vec{n}\wedge\partial^\nu\vec{n})].
\end{equation}
$\vec{\zeta}$ is an arbitrary vector field, therefore the equations of motion read
\begin{equation}\label{EOM}
\partial_\mu\vec{n}\;G^\mu=0
\end{equation}
with a magnetic current in natural units
\begin{equation}\label{magcurdimless}
G^\mu=\partial_\nu[\vec{n}(\partial^\nu\vec{n}\wedge\partial^\mu\vec{n})].
\end{equation}
This quadruple of equations defines magnetic currents and substitutes the homogeneous Maxwell equations. It reads in SI-units
\begin{equation}\label{magcur}
g^\mu=c\,\partial_\nu \hspace{0.2mm}{^*}\hspace{-0.2mm}f^{\nu \mu}
\quad \Leftrightarrow \quad
\begin{cases}
&\rho_\mathrm{mag} = \nabla \mathbf{B},\\
&\mathbf{g} = - \nabla \times \mathbf{E} - \partial_t\mathbf{B},
\end{cases}
\end{equation}
with
\begin{equation}
g^\mu=-\frac{e_0}{4\pi\varepsilon_0} G^\mu =(c\rho_{\rm mag},\mathbf{g}).
\end{equation}
The magnetic current is obviously conserved
\begin{equation}\label{conscurr}
\partial_\mu G^\mu=\partial_\mu \partial_\nu[\vec{n}(\partial^\nu\vec{n}\wedge\partial^\mu\vec{n})] = 0
\end{equation}
due to the antisymmetry of the expression in squared brackets of (\ref{conscurr}).

One can eliminate the colour degrees of freedom from the equations of motion (\ref{EOM}) by forming triple products
\begin{equation}\label{natEOM}
\vec{n} \vec{R}_{\mu \nu} G^\nu=
\vec{n} (\partial_\mu \vec{n} \wedge \partial_\nu\vec{n}) G^\nu=0.
\end{equation}
It reads in SI-quantities
\begin{equation}\label{obsEOM}
\hspace{0.8mm}{^*}\hspace{-0.8mm}f_{\mu \nu} g^\nu=0
\quad \Leftrightarrow \quad
\begin{cases}
&\mathbf{B} \; \mathbf{g} = 0,\\
&c^2 \mathbf{B} \; \rho_{\rm mag}  = \mathbf{g} \times \mathbf{E}.
\end{cases}
\end{equation}

We have therefore three quadruples of equations. The inhomogeneous Max\-well equations (\ref{inhmax}), the definition of magnetic currents (\ref{magcur}) and the equations of motion (\ref{obsEOM}). Note that only three of the Eqs.~(\ref{obsEOM}) are independent.

At the first glance the appearance of a magnetic current seems very unusual, but a closer look shows that this is a consequence of the topological restrictions of the $\vec{n}$-field. It is the price we have to pay for quantising the electric charges at the classical level. The $\vec{n}$-field is restricted to two degrees of freedom. The variation with respect to these two field does not suffice to get vanishing magnetic currents. It will be interesting to investigate whether the appearance of such a magnetic current leads to discrepancies with observations.

It should be mentioned that in this model the topological current is conserved \cite{Fab99}. This conservation gives the identity
\begin{equation}\label{identity}
 {\hspace{0.5mm}^*}\hspace{-0.5mm}\vec{R}_{\mu\nu}\vec{R}^{\mu\nu}=0.
\end{equation}
It follows that also in the electromagnetic limit electric and magnetic fields are orthogonal to each other
\begin{equation}\label{orthogonal}
\mathbf{E} \cdot \mathbf{B} = 0.
\end{equation}
This equation can also be derived at the level of $\vec{n}$-fields. We get
\begin{equation}
\begin{split}
\mathbf{E} \cdot \mathbf{B} & \propto
(\partial_0 \vec{n} \wedge \partial_1 \vec{n})(\partial_2 \vec{n} \wedge \partial_3 \vec{n}) +\\
&+(\partial_0 \vec{n} \wedge \partial_2 \vec{n})(\partial_3 \vec{n} \wedge \partial_1 \vec{n}) +\\
&+(\partial_0 \vec{n} \wedge \partial_3 \vec{n})(\partial_1 \vec{n} \wedge \partial_2 \vec{n}).
\end{split}
\end{equation}
Because of the normalisation $\vec{n}^2=1$ for non-vanishing electric or magnetic fields exactly two of the derivatives $\partial_\mu \vec{n}$ are linear independent. For simplicity we assume that $\partial_0 \vec{n}$ and $\partial_1 \vec{n}$ are linear independent. The other derivatives $\partial_2 \vec{n}$ and $\partial_3 \vec{n}$  can be written as linear combinations of $\partial_0 \vec{n}$ and $\partial_1 \vec{n}$
\begin{equation}\label{proofwithn}
\partial_2 \vec{n} = c_1 \partial_0 \vec{n} + c_2  \partial_1 \vec{n}, \quad
\partial_3 \vec{n} = c_3 \partial_0 \vec{n} + c_4  \partial_1 \vec{n},
\end{equation}
with coefficients $c_1$, $c_2$, $c_3$ and $c_4$. We get
\begin{equation}
\begin{split}
E_x B_x &\propto
(\partial_0 \vec{n} \wedge \partial_1 \vec{n})(\partial_2 \vec{n} \wedge \partial_3 \vec{n}) =\\
&=(c_1 c_4 - c_2 c_3)(\partial_0 \vec{n} \wedge \partial_1 \vec{n})^2,\\
E_y B_y &\propto
(\partial_0 \vec{n} \wedge \partial_2 \vec{n})(\partial_3 \vec{n} \wedge \partial_1 \vec{n}) =\\
&=c_2 c_3(\partial_0 \vec{n} \wedge \partial_1 \vec{n})^2,\\
E_z B_z &\propto
(\partial_0 \vec{n} \wedge \partial_3 \vec{n})(\partial_1 \vec{n} \wedge \partial_2 \vec{n}) =\\
&= - c_1 c_4(\partial_0 \vec{n} \wedge \partial_1 \vec{n})^2,\\
E_x B_x &+ E_y B_y + E_z B_z = 0.
\end{split}
\end{equation}
This agrees with Eq.~(\ref{orthogonal}), but it seems to contradict to the experimental situation. In many experiments we do not have orthogonal electric and magnetic fields, {\it e.g.} the presence of the magnetic field of the earth does not influence the electric field of a condenser. There is possibly a loophole in this argumentation. Eq.~(\ref{orthogonal}) is valid at the level of distances between elementary charges. At the level of measurable distances it could be possible that one has to average over the microscopic electric fields and over the microscopic magnetic fields and these averages need not be orthogonal. Eq.~(\ref{proofwithn}) shows clearly that the identity (\ref{orthogonal}) follows from the interpretation of the field strength as curvature of the unit vector field $\vec{n}$. It is not related to the Lagrangian and therefore to the dynamics of the fields. For vanishing magnetic currents the equation of motion (\ref{obsEOM}) would even allow for parallel $\mathbf{E}$ and $\mathbf{B}$.

From Eq.~(\ref{obsEOM}) follows that $\mathbf{B}$ is perpendicular to $\mathbf E$ and $\mathbf g$. $\mathbf E$ and $\mathbf g$ are in general not orthogonal to each other. But as we will discuss in the next section, for waves propagating freely in the vacuum $\mathbf E$ and $c\mathbf B$ have equal modulus. It follows from (\ref{obsEOM}) that the modulus of $\mathbf g$ equals the modulus of $g^0=c\rho_\mathrm{mag}$, {\it i.e.} $g^\mu$ is a light-like vector. The spatial relations between $\mathbf{B}$, $\mathbf E$ and $\mathbf g$ can be summarised in
\begin{equation}\label{aboutdirections}
\begin{split}
&\mathbf{B} \; \mathbf{g} = 0,\\
&\mathbf{B} \; \mathbf{E} = 0,\\
&\mathbf{E}\; \mathbf{g} + \frac{1}{\varepsilon_0} \mathbf{j}\mathbf{B} = c^2 \mathbf{B} (\nabla \times \mathbf{B}) - \mathbf{E} (\nabla \times \mathbf{E}).
\end{split}
\end{equation}

From the three quadruples (\ref{inhmax}), (\ref{magcur}) and (\ref{obsEOM})  one can further derive the relation
\begin{equation}\label{pointyng1}
\frac{1}{2} \partial_t \left( \varepsilon_0\mathbf{E}^2 
+ \frac{1}{\mu_0} \mathbf{B}^2 \right) 
=- \frac{1}{\mu_0} \nabla \left( \mathbf{E} \times \mathbf{B} \right) 
-\mathbf{ j}\mathbf{E}
\end{equation}
and the wave equations
\begin{eqnarray}\label{elwave}
&&\partial_\mu \partial^\mu \mathbf{E} 
= - \nabla \times \mathbf{g} -\mu_0 \partial_t \mathbf{j} -\frac{1}{\varepsilon_0} \nabla \rho,\\
&&\partial_\mu \partial^\mu \mathbf{B} 
= \mu_0 \nabla \times  \mathbf{j} - \frac{1}{c^2} \partial_t \mathbf{g} - \nabla \rho_{\rm mag}.\label{magwave}
\end{eqnarray}

These wave equations are related by a duality transformation in electric and magnetic quantities, but Eq.~(\ref{pointyng1}) is not. Eq.~(\ref{pointyng1}) is identical to the corresponding equation in Maxwell's electrodynamics, it describes the effects leading to a change of the energy density $\frac{1}{2} \partial_t \left( \varepsilon_0\mathbf{E}^2 + \frac{1}{\mu_0} \mathbf{B}^2 \right)$. The energy density decreases in the region of a source of the Poynting vector $\frac{1}{\mu_0} \nabla \left( \mathbf{E} \times \mathbf{B} \right)$ and in the region of electric charges moving along electric flux lines. Magnetic currents do not appear in Eq.~(\ref{pointyng1}). Since magnetic charges are only allowed to move perpendicular to magnetic flux lines, see Eq.~(\ref{obsEOM}), the energy density is not influenced by their presence.

Whereas electric charges are quantised, magnetic charges are not. Therefore, magnetic charges are not stable, they can be removed by smooth changes of the soliton field decreasing the corresponding contribution of the magnetic field to the energy density. We expect that in the ground-state magnetic charges are absent. Minimising the energy of the static electric field of a condenser leads to an electric field fulfilling $\nabla \times \mathbf{E}=0$, {\it i.e.} according to Eq.~(\ref{magcur}) to the absence of a magnetic current $\mathbf{g}$.

We can summarise that in the electromagnetic limit the $\vec{n}$-field defines on the same footing the electromagnetic field and its source $j^\mu(x)$ via its point-like singularities. These singularities correspond to quantised charges which propagate in time with velocities smaller than light only, see Eq.~(\ref{currvec}). The model of topological fermions differs from Maxwell's theory by the fact that it can only describe integer multiples of the elementary electric charge $e_0$ and by the appearance of magnetic currents. Since the magnetic field depends on the time derivative of the basic $\vec{n}$-field such magnetic currents are absent for static soliton configurations or solitons moving with constant velocity.

\section{Electromagnetic waves in the vacuum\label{Propagating}}

Now we will consider the case when the  $\vec{n}$-field has no singularity and thus $\rho=0$ and $\mathbf{j}=0$. Let us show that in this case electromagnetic waves, if any, should freely propagate in the vacuum with the speed of light. Explicit solutions for such waves will be discussed in a separate paper.

We will treat the propagation of a disturbance along (opposite to) the $z$-direction characterised by the relation
\begin{equation}\label{relderivs}
\partial_0 \vec{n} = - \beta \partial_z \vec{n} \quad \text{with} \quad \beta=\frac{v}{c}\,.
\end{equation}
From the beginning we have to remark that for vanishing magnetic current the wave equations (\ref{elwave}) and (\ref{magwave}) are reduced to the standard Maxwell equations and the condition $\beta^2=1$ follows directly from these wave equations. So we have to consider only the case of non-vanishing $g^\mu$.

From (\ref{relderivs}) we derive the following relations
\begin{equation}\label{magpropag}
\begin{split}
&cB_x=-\frac{e_0}{4\pi\varepsilon_0} \vec{n} \, \partial_0 \vec{n} \wedge \partial_x \vec{n}
=-\beta E_y,\\
&cB_y=-\frac{e_0}{4\pi\varepsilon_0} \vec{n} \, \partial_0 \vec{n} \wedge \partial_y \vec{n}
=\beta E_x,\\
&cB_z=-\frac{e_0}{4\pi\varepsilon_0} \vec{n} \, \partial_0 \vec{n} \wedge \partial_z \vec{n} = 0.\\
\end{split}
\end{equation}
The case of vanishing parallel electric field strength
\begin{equation}\label{vanelectric}
E_z = -\frac{e_0}{4\pi\varepsilon_0} \vec{n} \, \partial_x \vec{n} \wedge \partial_y \vec{n} = 0
\end{equation}
 we can interpret as the absence of charged matter leading to non-vanishing spatial curvature. Using (\ref{magcur}) we arrive at
\begin{equation}\label{gpropag}
\begin{split}
&g_0 = c \partial_x B_x + c \partial_y B_y 
= - \beta (\partial_x E_y - \partial_y E_x) = \beta g_z ,\\
&g_x = \partial_z E_y - \partial_t B_x
= (1-\beta^2) \partial_z E_y,\\
&g_y = - \partial_z E_x - \partial_t B_y
= (1-\beta^2) \partial_z E_x,\\
&g_z = - \partial_x E_y + \partial_y E_x.
\end{split}
\end{equation}
From the equations of motion (\ref{obsEOM}) follows
\begin{equation}\label{betaeq1}
\left.
\begin{aligned}
&c B_x g_0 + E_y g_z = (1-\beta^2) E_y g_z = 0\\
&c B_y g_0 - E_x g_z = -(1-\beta^2) E_x g_z = 0\\
\end{aligned}
\right\} \Rightarrow \beta^2=1,
\end{equation}
for non-vanishing $g^\mu$. The magnetic current $g^\mu$ turns out to be a light-like vector, $g^\mu = (\beta g_z, 0, 0, g_z)$,  with a $z$-component only which is  $g_z=-\mathrm{curl}\ \mathbf{E}$.

Plane waves have vanishing $\mathrm{curl}\ \mathbf{E}$ and therefore vanishing magnetic current. $g^\mu$ can only be non-zero at the boundary of electromagnetic rays. In this sense for waves propagating in the vacuum the appearance of magnetic currents would be a boundary effect.

In the presence of matter the parallel electric field strength (\ref{vanelectric}) may not vanish. This situation is more complicated, in electrodynamics it is usually described by introducing the relative dielectricity $\epsilon_r$ and the relative permeability $\mu_r$. The discussion of matter effects go beyond the scope of this article.

Note that owing to the structure of the MTF Lagrangian the time derivatives acting on the $\vec n$-field are multiplied by an induced metric on the colour manifold. Nevertheless this does not lead to effects similar to the Velo-Zwanziger phenomenon \cite{VZ69}, to acausal propagation of electromagnetic wave around the background of the electric monopoles. 

To show this it is sufficient to consider scattering waves in the wave zone. Scattering waves can be obtained expanding the Lagrangian up to second order in the fluctuations (\ref{vary}) around the hedgehog solution. The first order term, $\mathcal L^{(1)}$, vanishes for the solution of the equations of motion (\ref{EOM}). From the three components of the fluctuating field  $\vec\zeta(x)$ only two  modes are really independent (\ref{transv}). Neglecting terms which decay faster than $1/r^2$ the Lagrangian $\mathcal L^{(2)}$  reads (with $\epsilon=1$)
\begin{equation}\label{zetaLagrangian}
\mathcal{L}^{(2)}
 =\frac1{2r^2} \partial_\mu \vec\zeta \, \partial^\mu \vec\zeta
 + \mathcal{O}\left(\frac1{r^3}\right).
\end{equation}
The resulting field equation  
\begin{equation}\label{FieldEqsZeta}
\frac1{r^2} \partial_\mu\partial^\mu\vec\zeta(x) + \mathcal{O}\left(\frac1{r^3}\right)=0
\end{equation}
determines the fluctuating field $\vec \zeta(x)$. For the scattering wave its solution is
\begin{equation}\label{in-out-waves}
\vec\zeta(x)=\vec e \, f\left(t-\frac{r}{c}\right) +\mathcal{O}\left(\frac1{r}\right),
\end{equation}
where $f(t-r/c)$ is an arbitrary function and $\vec e$ is any unit vector orthogonal to the $\vec n$-vector defining the polarisation of the wave.

Of course the physical field is not $\vec \zeta(x)$, but the fluctuating field strength
\begin{equation}\label{FluctuatedWaves}
\begin{split}
\delta\,^*\hspace{-1mm}f_{\mu \nu}(x) = & - \frac{e_0}{4 \pi \varepsilon_0 c}\left(
\partial_\mu\vec \zeta  \wedge\partial_\nu \vec n +\partial_\mu\vec  n  \wedge \partial_\nu \vec \zeta
\right) \vec n + \\
& + \mathcal{O}\left(\frac1{r^2}\right).
\end{split}
\end{equation}
Substituting the solution (\ref{in-out-waves}) in (\ref{FluctuatedWaves}) one gets for the oscillating electric and magnetic fields 
\begin{equation}\label{outEMWaves}
\begin{split}
\delta\mathbf E &= - \, \mathbf e
 \, \frac{e_0}{4 \pi \varepsilon_0} \frac{f'\left(t-\frac{r}{c}\right)}r
 + \mathcal{O}\left(\frac1{r^2}\right), \\
\delta\mathbf B &= - \, (\mathbf e_r \times \mathbf e)
 \, \frac{e_0}{4 \pi \varepsilon_0 c} \frac{f'\left(t-\frac{r}{c}\right)}r
 + \mathcal{O}\left(\frac1{r^2}\right). 
\end{split}
\end{equation}
This is a usual outgoing electromagnetic wave radiated by the spherically symmetric soliton propagating with the velocity of light.

\section{Coulomb- and Lorentz forces\label{sec:Forces}}

In the MTF charges and their fields can't be separated \cite{Fab99}. The interaction of solitons is a consequence of topology \cite{HT93}. Forces between solitons and their fields are internal forces. The total force density does therefore vanish \cite{Fab99}. The electrodynamic limit gives a nice possibility to split solitons from their fields treating electric charges as external sources and to derive Coulomb and Lorentz forces. By the third law of Newton the force acting on charges is the counter force to that acting on fields.

 The basic field variable is the unit field $\vec{n}(x)$ and $\partial_\mu \vec{n}(x)$ are the generalised velocities. The energy-momentum tensor of the electromagnetic field is defined by the Lagrangian (\ref{EDLagrangian})
\begin{equation}\label{DefEMomTen}
\Theta^\mu_{\;\nu}(x) 
=\frac{\partial {\cal L_{\rm ED}}(x)}{\partial (\partial_\mu n_K)} \partial_\nu n_K - {\cal L_{\rm ED}}(x)\,\delta^\mu_\nu.
\end{equation}
It is symmetric already in its original form
\begin{equation}\label{ExpEMomTen}
\Theta^\mu_{\;\nu}(x) 
=- \frac{1}{\mu_0}{\hspace{0.5mm}^*}\hspace{-0.5mm}f_{\nu \sigma}(x){\hspace{0.5mm}^*}\hspace{-0.5mm}f^{\mu  \sigma}(x) - {\cal L_{\rm ED}}(x)\,\delta^\mu_\nu.
\end{equation}
For comparison we would like to mention that the canonical energy-momentum tensor in Maxwell's electrodynamics suffers from the lack of symmetry \cite{Ja75} and has to be especially symmetrised. The components of $\Theta^\mu_{\;\nu}$ read
\begin{eqnarray}\label{Theta00}
\Theta^0_{\;0} &=& {\cal H} = \frac{\varepsilon_0}{2} \left[ \mathbf{E}^2 + c^2\mathbf{B}^2 \right],\\
\Theta^0_{\;i} &=& - c \varepsilon_0 \, (\mathbf{E} \times \mathbf{B})_i,\\
\Theta^i_{\;j} &=&\varepsilon_0 \left[ E_i E_j + c^2 B_i B_j - \frac{\delta^i_j}{2} (\mathbf{E}^2 + c^2 \mathbf{B}^2) \right] \,  .
\end{eqnarray}
They have the same form as those for the symmetrised energy-momentum tensor in Maxwell's electrodynamics. As mentioned above the soliton model \cite{Fab99} is a closed system and therefore we get a zero total force density
\begin{equation}\label{totforczero}
\partial^\nu \Theta^\mu_{\;\nu} + f^\mu_\mathrm{charges} = 0
\end{equation}
consisting of the force density $\partial^\nu \Theta^\mu_{\;\nu}$ acting on the electromagnetic field and the force density $f^\mu_\mathrm{charges}$ acting on charges. In the electrodynamic limit charges appear as external sources. The force density acting on this sources we get therefore by
\begin{equation}\label{FD}
f^\mu_\mathrm{charges} = -\partial^\nu \Theta^\mu_{\;\nu}.
\end{equation}
Inserting the energy momentum tensor (\ref{ExpEMomTen}) in the r.h.s of this equation results in
\begin{widetext}
\begin{equation}\label{forcedensityderivation}
\begin{split}
f^\mu_\mathrm{charges}
&=\frac{1}{\mu_0} \partial^\nu \left( {\hspace{0.5mm}^*}\hspace{-0.5mm}
f_{\nu \rho} {\hspace{0.5mm}^*}\hspace{-0.5mm}f^{\mu \rho} \right)
-\frac{1}{4\mu_0}\partial^\mu({\hspace{0.5mm}^*}\hspace{-0.5mm} f_{\rho \nu}
{\hspace{0.5mm}^*}\hspace{-0.5mm} f^{\rho \nu})
=\frac{1}{\mu_0}
[ \underbrace{\partial^\nu {\hspace{0.5mm}^*}\hspace{-0.5mm} f_{\nu \rho }}_{\frac{1}{c} g_\rho} {\hspace{0.5mm}^*}\hspace{-0.5mm}f^{\mu \rho}
+\underbrace{{\hspace{0.5mm}^*}\hspace{-0.5mm}f_{\nu \rho}
\;\partial^\nu {\hspace{0.5mm}^*}\hspace{-0.5mm}f^{\mu \rho}}_{
-{\hspace{0.5mm}^*}\hspace{-0.5mm}f_{\nu \rho}
\partial^\rho {\hspace{0.5mm}^*}\hspace{-0.5mm}f^{\mu \nu}}
+\frac{1}{2}{\hspace{0.5mm}^*}\hspace{-0.5mm}f_{\nu \rho}
\;\partial^\mu {\hspace{0.5mm}^*}\hspace{-0.5mm}f^{\rho \nu} ]=\\
&=\frac{1}{\mu_0 c} \underbrace{
{\hspace{0.5mm}^*}\hspace{-0.5mm}f^{\mu \rho} g_\rho}_{0}
+\frac{1}{2\mu_0}{\hspace{0.5mm}^*}\hspace{-0.5mm}f_{\nu \rho}
[\underbrace{\partial^\rho {\hspace{0.5mm}^*}\hspace{-0.5mm}f^{\nu \mu}
+\partial^\nu {\hspace{0.5mm}^*}\hspace{-0.5mm}f^{\mu \rho}
+\partial^\mu {\hspace{0.5mm}^*}\hspace{-0.5mm}f^{\rho \nu}}_{
-\mu_0  \epsilon^{\mu \nu \rho \sigma} j_\sigma} ]
=-\frac{1}{2} \epsilon^{\mu \nu \rho \sigma}
{\hspace{0.5mm}^*}\hspace{-0.5mm}f_{\nu \rho} j_\sigma 
=f^{\mu \sigma} j_\sigma ,
\end{split}
\end{equation}
\end{widetext}
where we used in the under-braces the definition of magnetic currents (\ref{magcur}), the antisymmetry of ${\hspace{0.5mm}^*}\hspace{-0.5mm}f_{\nu \rho}$, the equations of motion (\ref{obsEOM}) and the definition of electric currents (\ref{inhmax}). Magnetic currents are only internal. Therefore according to the equations of motion (\ref{obsEOM}) they do not contribute directly to the force density acting on static and moving electric charges. More explicitly the force density (\ref{forcedensityderivation}) acting on charges reads
\begin{eqnarray}\label{forcedensity}
&&f^0_\mathrm{charges} = \frac{1}{c}\mathbf{j}\mathbf{E},\\
&&\mathbf{f}_\mathrm{charges} = \rho \mathbf{E} + \mathbf{j} \times \mathbf{B}.
\end{eqnarray}
The first of these equations is related to (\ref{pointyng1}) and describes the loss of power density of the field and the corresponding growth for particles. The second equation yields Coulomb and Lorentz forces acting on charged point-like solitons.

This shows that the electrodynamic limit of the soliton model corresponds to Maxwell's theory with a restriction of the possible field configurations to those of integer multiples of elementary charges. This has led also to non-vanishing internal magnetic currents which do not influence directly the motion of charged particles.

In electrodynamics the gauge field $A^\mu(x)$ is considered as the basic field and defines the set of possible field configurations. In the vacuum two of the components of $A^\mu$ turn out to be unnecessary and can be gauged away. The remaining two degrees of freedom are sufficient to describe the two orthogonal polarisations of electromagnetic waves. In the following section we will derive that in the electrodynamic limit of the MTF there appears a U(1)-gauge symmetry.

\section{U(1) gauge invariance\label{sec:gauge}}

There is a description of electrodynamics which uses the dual potential $c^\mu(x)$ \cite{BBZ94}. It introduces electric charges via Dirac strings which are characterised by singularities of the gauge field $c^\mu$. This formulation of electrodynamics is closely related to the model of topological fermions where the singularities of the connection field are removed via the additional degree of freedom $\alpha(x)$ of the soliton field $Q(x)= \cos\alpha(x)+i\sin\alpha(x) \vec{\sigma}\vec{n}$. In the soliton description one can clearly see the relation of the gauge freedom of electrodynamics to geometry. We will now have a closer look at this relation. In the electrodynamic limit the basic field is the $\vec{n}$-field. The connection $\vec{\Gamma}_\mu(x)=\partial_\mu \vec{n}\wedge\vec{n}$ turns out to be the ``angular frequency'' of $\vec{n}(x)$ in the direction $\delta x^\mu$
\begin{equation}\label{angfreq}
\begin{split}
e^{i{\vec\Gamma}_\mu(x)\vec{T} \delta x^\mu} \vec{n}(x)&=
\vec{n}+i \delta x^\mu(\vec{\Gamma}_\mu\vec{T})\vec{n}=\\
&=\vec{n}-(\partial_\mu \vec{n}\wedge\vec{n})\wedge\vec{n}\;\delta x^\mu=\\
&=\vec{n}+\partial_\mu \vec{n}\;\delta x^\mu=\vec{n}(x+\delta x).
\end{split}
\end{equation}
There is a simple geometrical relation of the curvature $\partial_{\mu}\vec{n}\wedge\partial_{\nu}\vec{n}$ and the curl of the connection field which can be read off from the tangential component of $\partial_\mu \partial_\nu \vec{n}$ using its symmetry under exchange of $\mu$ and $\nu$
\begin{equation}\label{u1maucar}
\begin{split}
\vec{n}\wedge\partial_\mu \partial_\nu\vec{n}&=
\partial_\mu(\vec{n}\wedge\partial_\nu\vec{n})-\partial_{\mu}\vec{n}\wedge\partial_{\nu}\vec{n}=\\
&=\partial_\nu(\vec{n}\wedge\partial_\mu\vec{n})+\partial_{\mu}\vec{n}\wedge\partial_{\nu}\vec{n}.
\end{split}
\end{equation}
This is a special case of the Maurer-Cartan equation (\ref{maurercartan}). The relations (\ref{angfreq}) and (\ref{u1maucar}) have to be understood in terms of the soliton field $i\vec{\sigma}\vec{n}$, where $\vec{n}$ is multiplied with the quaternionic units $-i\vec{\sigma}$. They are therefore purely geometrical and independent of the basis system. If the quaternionic units are rotated by some orthogonal matrix field $\Omega(x)$
\begin{equation}\label{sigtrans}
\sigma^\prime_A=\Omega_{AB}\sigma_B
\end{equation}
with
\begin{equation}
\begin{split}
\Omega(x)&=e^{-i\vec{\omega}(x)\vec{T}}=e^{-i\omega\vec{e}_\omega\vec{T}} =e^{-i\omega T_\omega} =\\
&=1-i \sin \omega T_\omega + (\cos \omega - 1) T_\omega^2 
\end{split}
\end{equation}
also $\vec{n}$ has to transform to $\vec{n}^{\,\prime}$ under such a gauge transformation
\begin{equation}\label{ntrans}
\begin{split}
\vec{n}^{\,\prime}&=\Omega\vec{n}=\\
&=(\vec{n}\vec{e}_\omega)\vec{e}_\omega+\sin\omega\;\vec{e}_\omega\wedge\vec{n}-\cos\omega\;\vec{e}_\omega\wedge(\vec{e}_\omega\wedge\vec{n}),
\end{split}
\end{equation}
where the component of $\vec{n}$ parallel to $\vec{e}_\omega$ remains invariant and the perpendicular component is rotated by $\omega$. From $\partial_\mu\vec{n}=i\Gamma_\mu\vec{n}$ with $\Gamma_\mu=\vec{\Gamma}_\mu\vec{T}$ follows
\begin{equation}\label{nderivtrans}
\partial_\mu\vec{n}^{\,\prime}=i\Gamma_\mu^\prime\vec{n}{\,^\prime}=
-\vec{\Gamma}_\mu^\prime\wedge\vec{n}^{\,\prime}, \quad
\Gamma_\mu^\prime=\Omega(\Gamma_\mu+i\partial_\mu)\Omega^\dagger
\end{equation}
and
\begin{eqnarray}\label{nGammatrans}
\vec{\Gamma}_\mu^\prime&=&\Omega\vec{\Gamma}_\mu+\vec{\Omega}_\mu, \nonumber\\
\vec{\Omega}_\mu&=&\frac{1}{2i} {\rm tr} (\vec{T}\;\partial_\mu\Omega\;\Omega^\dagger)=\\
&=&-\partial_\mu\omega\;\vec{e}_\omega-\sin\omega\;\partial_\mu\vec{e}_\omega-(\cos\omega-1)(\vec{e}_\omega\wedge\partial_\mu\vec{e}_\omega). \nonumber
\end{eqnarray}

The covariant derivative $D_\mu=\partial_\mu-i\Gamma_\mu$ has the important property that acting on an arbitrary vector field $\vec{v}(x)$ it is rotated by $\Omega$ only
\begin{equation}\label{covder}
D_\mu^\prime\Omega\vec{v}=(\partial_\mu-i\Gamma_\mu^\prime)\Omega\vec{v}=
\Omega(\partial_\mu-i\Gamma_\mu)\vec{v}.
\end{equation}
We can express the curvature tensor $\vec{R}_{\mu\nu}$ by the covariant derivative
\begin{equation}\label{curvat}
\vec{R}_{\mu\nu}\vec{T}=-i(D_\mu D_\nu - D_\nu D_\mu).
\end{equation}
It is obvious that this expression for $\vec{R}_{\mu\nu}$ is also gauge covariant
\begin{equation}\label{gaugecovR}
\vec{R}_{\mu\nu}^\prime\vec{T}=-i(D_\mu^\prime D_\nu^\prime - D_\nu^\prime D_\mu^\prime)=\Omega\vec{R}_{\mu\nu}\vec{T}\Omega^\dagger
\end{equation}
with
\begin{equation}\label{generalR}
\vec{R}_{\mu\nu}^\prime=
\partial_\nu\vec{\Gamma}_\mu^\prime-\partial_\mu\vec{\Gamma}_\nu^\prime
+\vec{\Gamma}_\nu^\prime\wedge\vec{\Gamma}_\mu^\prime
\end{equation}
which agrees with (\ref{intcurvature1}) in the original coordinate system.

In the rotated coordinate system the connection $\vec{\Gamma}_{\mu}^\prime$ may differ from $\partial_{\mu}\vec{n}^\prime\wedge \vec{n}^\prime$. According to (\ref{nderivtrans}) it may have a component in the direction of $\vec{n}^{\,\prime}$
\begin{equation}\label{nprimecomp}
\vec{\Gamma}_\mu^\prime=
\partial_{\mu}\vec{n}^{\,\prime}\wedge \vec{n}^{\,\prime}+\vec{n}^{\,\prime}(\vec{n}{\,^\prime}\vec{\Gamma}_\mu^\prime).
\end{equation}
Using (\ref{nGammatrans}) this component reads
\begin{equation}\label{parcomp}
\vec{n}^{\,\prime}\vec{\Gamma}_\mu^\prime=
\vec{n}^{\,\prime}\Omega\vec{\Gamma}_\mu+\vec{n}^{\,\prime}\vec{\Omega}_\mu=
\vec{n}\vec{\Gamma}_\mu+\vec{n}^{\,\prime}\vec{\Omega}_\mu=
\vec{n}^{\,\prime}\vec{\Omega}_\mu.
\end{equation}

There is a special type of transformations which leave $\vec{n}$ and $\vec{R}$ unchanged, whereas $\vec{\Gamma}_\mu$ varies, $\vec{\Gamma}_\mu\to \vec{\Gamma}_\mu^\prime$. These are the transformations with $\vec{e}_\omega=\vec{n}$ which according to (\ref{nprimecomp}), (\ref{parcomp}) and (\ref{nGammatrans}) lead to
\begin{equation}\label{Gamprime}
\vec{\Gamma}_\mu^\prime=
\partial_\mu\vec{n}\wedge\vec{n}-\partial_\mu\omega\;\vec{n}.
\end{equation}
These rotations around the $\vec{n}$-axis correspond to the Abelian gauge symmetry of electrodynamics. After such rotations $\vec{\Gamma}_\mu^\prime$ does not satisfy the Maurer-Cartan equations (\ref{u1maucar}). Inserted in expression (\ref{generalR}) for the curvature tensor the contributions of $\partial_\mu \omega$ cancel and give the gauge-invariant expression $\vec{R}_{\mu\nu}^\prime=\vec{R}_{\mu\nu}=\partial_{\mu}\vec{n}\wedge\partial_{\nu}\vec{n}$.

The alignment of $\vec{n}^\prime$ in 3-direction via
\begin{equation}\label{gotoabel}
\begin{split}
\Omega(x)=e^{-i\theta(x)\vec{e}_\phi(x)\vec{T}}, \qquad
\vec{n}^\prime=\Omega\vec{n}=
\left( \begin{array}{c}0\\0\\1\\\end{array} \right)=\vec{e}_3,\\
\omega=-\theta, \qquad 
\vec{e}_\omega=\vec{e}_\phi=\cos\phi\;\vec{e}_2-\sin\phi\;\vec{e}_1,
\end{split}
\end{equation}
where $\vec{e}_\phi$ is the unit vector in spherical coordinates in internal $\phi$-direction, leads to the Abelian Maxwell-theory in dual formulation. From (\ref{nprimecomp}) follows that $\vec{\Gamma}_\mu^\prime$ is also aligned in 3-direction. With (\ref{nprimecomp}), (\ref{parcomp}),(\ref{nGammatrans}),  and $\partial_\mu\vec{e}_\phi=-\partial_\mu\phi\,(\cos\phi\,\vec{e}_1+\sin\phi\,\vec{e}_2)$ we get
\begin{equation}\label{Gammaabel}
\begin{split}
\vec{\Gamma}_\mu^\prime&=
\vec{e}_3(\vec{e}_3\vec{\Omega}_\mu)=
(\cos\theta-1)\,\vec{e}_3[\vec{e}_3\cdot\vec{e}_\phi\wedge\partial_\mu\vec{e}_\phi]=\\
&=(\cos\theta-1)\partial_\mu\phi\;\vec{e}_3.
\end{split}
\end{equation}
The curvature (\ref{generalR}) contains only the curl term
\begin{equation}\label{colourRabel}
\vec{R}_{\mu\nu}^\prime=
\partial_\nu\vec{\Gamma}_\mu^\prime-\partial_\mu\vec{\Gamma}_\nu^\prime=
(\partial_\mu\theta\,\partial_\nu\phi-\partial_\nu\theta\,\partial_\mu\phi)\sin\theta\,\vec{e}_3
\end{equation}
and points also in 3-direction. This gives a general relation between the Abelian field strength $f_{\mu \nu}(x)$ (\ref{AbelianFS}) and the soliton field $\vec{n}(x)$
\begin{equation}\label{Rabel}
\begin{split}
R_{\mu \nu}(x) &= \vec{R}_{\mu\nu}^\prime \,\vec{e}_3
= \vec{R}_{\mu\nu} \vec{n} =\\
&=- \partial_\mu \cos \theta\,\partial_\nu\phi
+ \partial_\nu \cos \theta\,\partial_\mu\phi
\end{split}
\end{equation}
which can also be derived from the spherical representation of $\vec{n}(x)$ and $\vec{R}_{\mu\nu} \vec{n}=\partial_\mu \vec{n} \wedge \partial_\nu \vec{n}$.

A special case is the hedgehog-solution $\vec{n}=\frac{\mathbf{r}}{r}$ which identifies the internal coordinates $\theta$ and $\phi$ with the spherical coordinates in physical space $\vartheta$ and $\varphi$. Only one of the components of the curvature tensor, $\vec{R}_{\vartheta\varphi}^\prime$, is non-vanishing. We get from (\ref{Rabel})
\begin{equation}
\vec{R}_{\vartheta\varphi}^\prime=\sin\theta\,\vec{e}_3.
\end{equation}
The expression for the corresponding component of the field-strength tensor, $\vec{E}_{r}^\prime$, follows after dividing by the corresponding area element $r^2\sin\vartheta$ and multiplying with a measure system dependent factor
\begin{equation}
\vec{E}_{r}^\prime=-\frac{e_0}{4\pi\varepsilon_0}\frac{\vec{R}_{\vartheta\varphi}^\prime}{r^2\sin\vartheta}=-\frac{e_0}{4\pi\varepsilon_0}\frac{\vec{e}_3}{r^2}.
\end{equation}

After the rotation (\ref{gotoabel}) of $\vec{n}$ in 3-direction there is a $U(1)$-symmetry left. This is due to possible rotations around the 3-direction which modify the connection $\vec{\Gamma}_\nu^\prime$ but not the curvature term $\vec{R}_{\mu\nu}^\prime$. Since it is not possible to comb a sphere, in the Abelian description the point-like singularities of the soliton-model are connected by line-like singularities, the well-known Dirac-strings \cite{Di48}. This derivation demonstrates that in the electrodynamic limit the model of topological fermions reduces to a dual formulation of Dirac's extension of electrodynamics with magnetic currents restricted by Eq.~(\ref{obsEOM}).

\section{Homogeneous electric fields\label{sec:homogeneous}}

A constant electric field in $z$-direction, $E_i=E_z \delta_{iz}$, fulfils the standard Maxwell-equations. In the electrodynamic limit the basic field is the unit-vector field $\vec n(x)$. A constant $\vec n$-field corresponds to zero field strength. Due to the topological restriction one can get a homogeneous electric field only in a finite spatial region. To describe this field let us introduce cylindrical coordinates, $z$, $r_\perp$ and $\varphi$, assume an axial symmetric problem with a profile function $E_z(r_\perp)$ with $N/2$ electric flux units $\frac{e_0}{\varepsilon_0}$
\begin{equation}\label{EFluss}
2\pi\int_0^\infty dr_\perp r_\perp  E_z(r_\perp) = \frac{N}{2}\frac{e_0}{\varepsilon_0}.
\end{equation}
To get the corresponding $\vec n$-field for a homogeneous electric field it is convenient to start from Eq.~(\ref{Rabel}). Then Eqs.~(\ref{dualfield}) and (\ref{AbelianFS}) lead to
\begin{eqnarray}\label{EzVonr}
E_z &:=& - \frac{e_0}{4\pi \varepsilon_0} R_{xy},\\
R_{xy} &:=& \vec{R}_{xy}\vec{n}
= \vec{R}_{xy}^\prime \vec{e}_3
=-\partial_x\cos \theta\,\partial_y\phi+\partial_y\cos \theta\,\partial_x\phi, \nonumber
\end{eqnarray}
relating the electric field strength $E_z$ to the direction of the $\vec{n}$-field
\begin{equation}\label{nvonthetaundphi}
\vec{n}=(\sin \theta \cos \phi, \sin \theta \sin \phi, \cos \theta)
\end{equation}
in colour space. A reasonable ansatz for the coordinate dependence of the colour angles $\theta$ and $\phi$ is
\begin{equation}\label{Nphi}
\phi= N \varphi, \quad \theta=\theta(r_\perp).
\end{equation}
The following boundary condition 
\begin{equation}\label{BC_homoge}
\theta(0)=0
\end{equation}
defines a global gauge fixing of the internal coordinate system.

\begin{figure}
  \psfrag{x}{$x$}
  \psfrag{y}{$y$}
  \psfrag{z}{$z$}
  \psfrag{f}{$\varphi$}
  \centering
  \includegraphics[height=0.2\textheight]{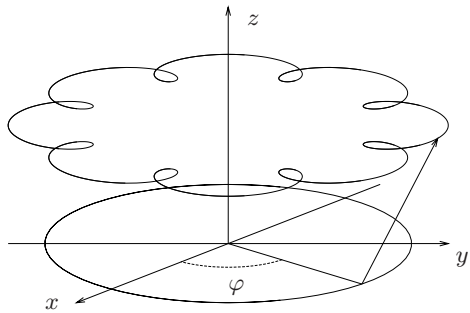}
  \caption{Behaviour of the vector $\vec n$. One rotation around the $z$-axis in physical space generates $N$ rotations of the vector $\vec n$ around
the 3-axis in colour space.}
  \label{nrotations}
\end{figure}

With
\begin{equation}
\begin{split}
&\partial_x \cos \theta 
= \frac{\partial r_\perp}{\partial x} \, \partial_\perp \cos \theta 
= \cos \varphi \; \partial_\perp \cos \theta,\\
&\partial_y \cos \theta 
= \frac{\partial r_\perp}{\partial y} \, \partial_\perp \cos \theta 
= \sin \varphi \; \partial_\perp \cos \theta,\\
&\partial_x \phi = N \, \partial_x \varphi
= -N \frac{\sin \varphi}{r_\perp},\\
&\partial_y \phi = N \, \partial_y \varphi
= N \frac{\cos \varphi}{r_\perp}, \quad
\end{split}
\end{equation}
we get from (\ref{EzVonr}) the differential equation
\begin{equation}\label{RxyVonr}
\begin{split}
R_{xy}(r_\perp)&=-\partial_x\cos \theta\,\partial_y\phi+\partial_y\cos \theta\,\partial_x\phi=\\
&=-\frac{N}{r_\perp} \partial_\perp \cos \theta(r_\perp).
\end{split}
\end{equation}
With the boundary condition (\ref{BC_homoge}) it follows from (\ref{RxyVonr}) and (\ref{EzVonr})
\begin{equation}\label{CosTheta}
\begin{split}
\cos \theta(r_\perp) &= - \frac{1}{N} \int_0^{r_\perp} d\rho \rho R_{xy}(\rho)=\\
&= \frac{4\pi \varepsilon_0}{N e_0} \int_0^{r_\perp} d\rho \rho E_z(\rho)
\end{split}
\end{equation}
which for $\theta(\infty)=\pi$ is in agreement with the flux condition (\ref{EFluss}). In Eq.~(\ref{EFluss}) there appear only $\frac{N}{2}$ flux quanta, since half of the flux of $N$ elementary charges on a condenser plate is going in positive $z$- and half in negative $z$-direction. In the central region of constant field strength $E_z(\rho)$  the $r_\perp$-dependence of $\cos \theta(r_\perp)$ reads
\begin{equation}\label{constfield}
\cos \theta(r_\perp) = \frac{2 \varepsilon_0}{N e_0 } r_\perp^2 \pi E_z(0).
\end{equation}
This is the electric flux through $r_\perp^2\pi$ divided by the total number of flux quanta. Eq.~($\ref{constfield}$) defines together with Eq.~(\ref{Nphi}) the coordinate dependence of the colour vector $\vec{n}(z,r_\perp,\varphi)$ in Eq.~(\ref{nvonthetaundphi}). A schematic picture of the behaviour of $\vec{n}$ is depicted in Fig.~\ref{nrotations}.

This solution (\ref{nvonthetaundphi}),(\ref{Nphi}),(\ref{CosTheta}) looks more complicated than that in Maxwell's theory, but it agrees better with the experimental situation with condenser plates of finite size and finite charge $Ne_0$ producing a field whose homogeneity is only approximate.

\section{Conclusion\label{sec:Summary}}

In the MTF the vacuum states are degenerate on an internal $\mathrm{S}^2_\mathrm{equ}$-sphere. Therefore, there exists a limit, the electrodynamic limit $q_0=0$, $Q=i\vec{\sigma}\vec{n}$, far away from the soliton centres, where the system is in the vacuum state of broken symmetry. This limit corresponds to the situation when the length scale parameter $r_0$ is sent to zero and solitons shrink to dual Dirac monopoles. The remaining two degrees of freedom are sufficient to describe the electromagnetic field. The Lagrangian of this model reduces to the Lagrangian of the Maxwell field in dual representation.

We have derived that in the electrodynamic limit the system obeys the inhomogeneous Maxwell-equations. Due to the topological restrictions electromagnetic waves may be accompanied by magnetic currents which in the vacuum propagate with the velocity of light in the direction given by the Poynting vector $\mathbf{E} \times \mathbf{B}/\mu_0$. Solitons behave as electrons with charges originating in their topological structure. The interaction of these charges via Coulomb- and Lorentz-forces turns out to be a consequence of topology.
A further consequence of topology is the prediction of the orthogonality of electric and magnetic fields at a microscopic level. This is some problem in comparison to experiment which may have a possible resolution in the fact that averages of fields over measurable distances need not be orthogonal.

The U(1) gauge invariance of electrodynamics can be derived from geometry, especially from the U(1) rotational invariance around the $\vec{n}$ axis. Finally we applied a general relation between the Abelian field strength $f_{\mu \nu}(x)$ and the $\vec{n}$-field to represent homogeneous fields in the MTF. It is interesting to see that only integer charges and homogeneous fields with quantised flux can be described. The set of fields is restricted by topology. Integer multiples of the elementary charges $e_0$ and fields homogeneous in some restricted area only can be represented.

\vspace{5mm}
\begin{acknowledgments}
We wish to acknowledge interesting discussions with Andrei Ivanov and Dima Borisyuk.
\end{acknowledgments}

\appendix

\section{ Relation of fields to geometry}\label{Geometry}

\begin{figure*}[h]
\centering
  \psfrag{R3}{\textcolor{Cyan4}{$\mathrm{R}^3$}}
  \psfrag{S3}{\textcolor{Cyan4}{$\mathrm{S}^3$}}
  \psfrag{S2}{\textcolor{Magenta4}{$\mathrm{S}^2_\mathrm{equ}$}}
  \psfrag{x}{$x$}
  \psfrag{y}{$y$}
  \psfrag{z}{$z$}
  \psfrag{dx}{\textcolor{Red2}{$dx$}}
  \psfrag{dy}{\textcolor{Red2}{$dy$}}
  \psfrag{U}{\textcolor{Cyan4}{$Q$}}
  \psfrag{u0}{\textcolor{Blue2}{$q_0$}}
  \psfrag{u1}{$q_1$}
  \psfrag{u2}{$q_2,q_3$}
  \psfrag{Gx}{\textcolor{Red2}{$\vec{\Gamma}_x dx$}}
  \psfrag{Gy}{\textcolor{Red2}{$\vec{\Gamma}_y dy$}}
  \psfrag{Gxy}{\textcolor{LtBlue}{$\vec{\Gamma}_x \wedge \vec{\Gamma}_y dx dy$}}
  \includegraphics[width=0.9\textwidth]{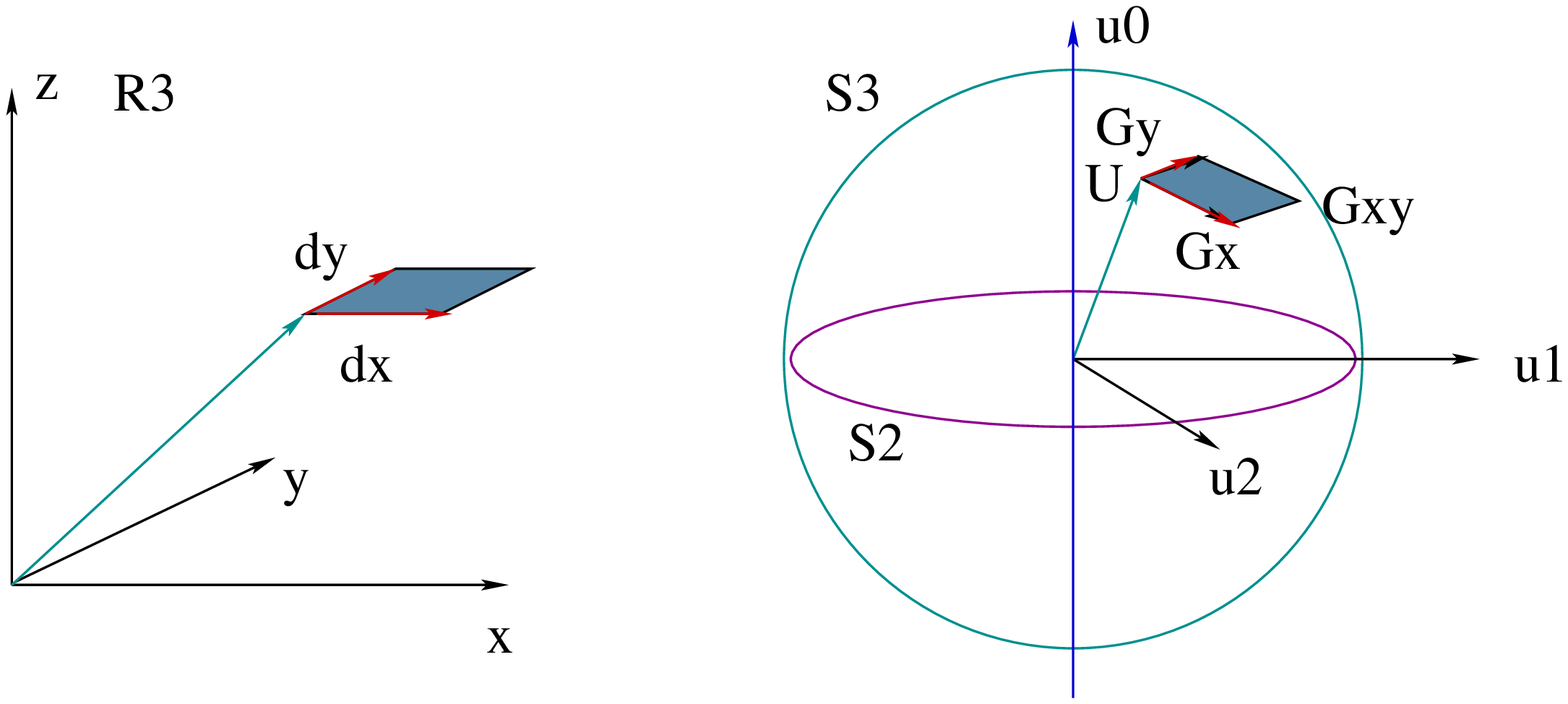}
  \caption{Map from a rectangle at $\vec{r}$ in configuration space to a parallelogram in a tangential plane of the internal ``colour'' space $\mathrm{S}^3$.}
  \label{map}
\end{figure*}
In the MTF the (dual) vector field $\vec{C}^\mu$ and the dual field strength $\hspace{0.8mm}{^*}\hspace{-0.8mm}\vec{F}^{\mu \nu}$ get a strong relation to geometry. The dual vector field is proportional to the connection field, the tangential vector $\vec{\Gamma}^\mu$ from $Q(x)$ to $Q(x+dx^\mu)$, see  Fig.~\ref{map}. Using $i \sigma_K Q$ as basis vectors of the tangential space at $Q$ we define
\begin{equation}\label{dualgaugefield}
\vec{C}^\mu = - \frac{e_0}{4 \pi \varepsilon_0} \vec{\Gamma}^\mu, \quad \partial^\mu Q = i \vec{\Gamma}^\mu \vec{\sigma} Q.
\end{equation}
The electric flux through a rectangle in configuration space is given by the corresponding area in the tangential plane of $\mathrm{S}^3$. This suggests to identify the dual field strength tensor with the curvature tensor $\vec{R}^{\mu \nu}$
\begin{equation}
\hspace{0.8mm}{^*}\hspace{-0.8mm}\vec{F}^{\mu \nu} = - \frac{e_0}{4 \pi \varepsilon_0 c} \vec{R}^{\mu \nu}, \quad \vec{R}^{\mu \nu} = \vec{\Gamma}^\mu \wedge \vec{\Gamma}^\nu.
\end{equation}

The general connection field $\vec{\Gamma}^\mu$ in (\ref{dualgaugefield}) is based on the soliton field $Q(x)$ at $x$. Moving along an infinitesimal rectangle in space-time, see Fig.~\ref{map}, leads back to the original soliton field $Q(x)$. From that follows that the connection field $\vec{\Gamma}^\mu$ obeys the Maurer-Cartan equation \cite{Fab99}
\begin{equation}\label{maurercartan}
\vec{\Gamma}^\mu \wedge \vec{\Gamma}^\nu = \frac{1}{2} \left( \partial^\nu \vec{\Gamma}^\mu - \partial^\mu \vec{\Gamma}^\nu \right).
\end{equation}
The curvature tensor can therefore be represented in a form which is well known from non-Abelian gauge theories
\begin{equation}\label{intcurvature}
\vec{R}^{\mu \nu} = \partial^\nu \vec{\Gamma}^\mu - \partial^\mu \vec{\Gamma}^\nu - \vec{\Gamma}^\mu \wedge \vec{\Gamma}^\nu.
\end{equation}
In those theories the gauge field $\vec{\Gamma}^\mu$ is the basic field and can't be derived from a soliton field.



%

\begin{thebibliography}{9}
%
\bibitem{Skyrme61} T.H.R.~Skyrme, Proc. Roy. Soc., {\bf A260} (1961) 127.
%
\bibitem{Witten} E.~Witten, Nucl. Phys. {\bf B 160} (1979) 57; ibid {B 223} (1983) 433.
%
\bibitem{ANW} G.S.~Adkins, C.R.~Nappi, E.~Witten, Nucl. Phys. {\bf B 228} (1983) 552.
%
\bibitem{MRS} V.G.~Makhankov, Y.P.Rybakov and V.I.~Sanyuk, The Skyrme Model. Springer-Verlag, 1993.
%
\bibitem{Fab99} M.~Faber, Few Body System 30 (2001) 149, hep-th/9910221.
%
\bibitem{HD64}
R.H.~Hobart, Proc. Phys. Soc. London, {\bf 82} (1963) 201; G.H.~Derrick, J. Math. Phys., {\bf 5} (1964) 1252.
%
\bibitem{Di48} P.A.M.~Dirac, Phys. Rev., {\bf 74} (1948) 817.
%
\bibitem{Raja} R.~Rajaraman, Solitons and Instantons, North-Holland, 1982.
%
\bibitem{VZ69}
G.~Velo and D.~Zwanziger, Phys. Rev., {\bf 186} (1969) 1337.
%
\bibitem{HT93} C.~Hong--Mo, T.~Sheung Tsun, Some Elementary Gauge Theory Concepts, World Scientific Lecture Notes in Physics - Vol 47, World Scientific, Singapore, 1993.
%
\bibitem{Ja75} J.D.~Jackson, Classical Electrodynamics, 
John Wiley $\&$ Sons, Inc., New York, Second Edition, 1975.
%
\bibitem{BBZ94} M.~Baker, J.~S. Ball, and F.~Zachariasen, ``Classical electrodynamics with dual
  potentials,'' \href{http://xxx.lanl.gov/abs/hep-th/9403169}{{\tt
  hep-th/9403169}}.
%
\end{thebibliography}
\end{document}